\documentclass[manuscript]{aastex}

\def\kms{$\rm km~s^{-1}$ }

\def\C3{$\rm C~III$}
\def\OO5{$\rm O~V$}
\def\N3{$\rm N~III$ }
\def\O6{$\rm O~VI$}
\def\fe18{$\rm [Fe~XVIII]$}
\def\Si12{$\rm Si~XII$}
\def\Al11{$\rm Al~XI$}
\def\si8{$\rm [Si~VIII]$}
\def\Fe10{$\rm [Fe~X]$}
\def\Ni14{$\rm [Ni~XIV]$}
\def\Ca14{$\rm [Ca~XIV]$}
\def\SSi3{$\rm Si~III$ }

\begin{document}

\title{Grain Destruction in a Supernova Remnant Shock Wave}
\shorttitle{Grain Destruction in an SNR Shock Wave}

\author{John C. Raymond\altaffilmark{1}} 
\author{Parviz Ghavamian\altaffilmark{2}}
\author{Brian J. Williams\altaffilmark{3}}
\author{William P. Blair,\altaffilmark{4} }
\author{Kazimierz J. Borkowski\altaffilmark{5}}
\author{Terrance J. Gaetz\altaffilmark{1}}
       \and
\author{Ravi Sankrit\altaffilmark{6}}

\altaffiltext{1}{Harvard-Smithsonian Center for Astrophysics, 60 Garden St., 
Cambridge, MA  02138, USA; jraymond@cfa.harvard.edu}
\altaffiltext{2}{Dept. of Physics, Astronomy \& Geosciences, Towson University, Towson, MD  21252}
\altaffiltext{3}{NASA Goddard Space Flight Center, Greenbelt, MD  20771}
\altaffiltext{4}{Department of Physics and Astronomy, Johns Hopkins University, 3400 N. Charles St., Baltimore, MD 21218, USA}
\altaffiltext{5}{Department of Physics, North Carolina State University, Raleigh, NC  27695}
\altaffiltext{6}{SOFIA Science Center, NASA Ames Research Center, M/S 232-12, Moffett Field, CA 94035}

\begin{abstract}
Dust grains are sputtered away in the hot gas behind
shock fronts in supernova remnants, gradually enriching the
gas phase with refractory elements.
We have measured emission in C IV $\lambda$1550 from C atoms
sputtered from dust in the gas behind a non-radiative shock
wave in the northern Cygnus Loop.  Overall, the intensity observed 
behind the shock agrees
approximately with predictions from model calculations that match
the {\it Spitzer} 24 $\mu$m and the X-ray intensity profiles.  
Thus these observations confirm the overall picture of dust destruction in SNR shocks
and the sputtering rates used in models.  However, there is a discrepancy in that the
CIV intensity 10$^\prime$$^\prime$ behind the shock is too high compared to the
intensities at the shock and 25$^\prime$$^\prime$ behind it.  
Variations in the density, hydrogen neutral fraction and the dust properties over 
parsec scales in the pre-shock medium limit our ability to 
test dust destruction models in detail.

\end{abstract}


\section{Introduction}

Destruction of dust by supernova remnant (SNR) shock waves controls the dust/gas ratio
in the ISM \citep{draine09, dwekscalo, dwek98}.  In SNRs, it also controls the gas phase abundances
of refractory elements such as C, Si and Fe, which are highly depleted in the
pre-shock gas but contribute strongly to SNR X-ray spectra.  It also determines the 
infrared cooling rate, which dominates over the X-ray cooling rate in shocks faster than 
about 400 \kms \citep{arendt10}.
However, the destruction rate is poorly known \citep{nozawa06},
and different types of grains are destroyed at different rates \citep{serra08}.

\medskip
In radiative shock waves, in which the density increases as the gas cools, 
grains collide with each other at high speed due to betatron acceleration.
The colliding grains shatter, altering the size distribution \citep{shull78, 
borkowskidwek, jones96, slavin04}.
In non-radiative shocks, for which the cooling time is large compared to dynamical 
time scales, sputtering dominates over grain shattering 
\citep{borkowski06}.  SNR shocks faster than 300 $\rm km~s^{-1}$ are typically
non-radiative \citep{cox}.  In the shocked plasma, proton and He$^{++}$ collisions
sputter atoms off the grains \citep{ds79}.  The
grains initially move at 3/4 of the shock speed relative to the plasma, and they 
gradually slow down due to gas drag and Coulomb drag.
For a shock moving perpendicular to the magnetic field (quasi-perpendicular shock),
 the motion is gyrotropic, 
while for a quasi-parallel shock the motion is along the flow direction.  Until a
grain slows down due to gas drag, its sputtering rate is enhanced by the increased
collision speed.  This nonthermal sputtering is especially important at moderate shock
speeds up to about 400 \kms \citep{dwek96}, typical of middle-aged SNRs such as
the Cygnus Loop.  Laboratory studies and computer simulations give sputtering rates 
\citep{bs05}, but the
simulated surfaces may differ from actual interstellar grains, and therefore the rates 
are uncertain.

\medskip
{\it Spitzer} observations have provided important new constraints on the mass, 
size distribution, temperature distribution and destruction rates of ISM 
grains in SNRs.  \citet{borkowski06} and \citet{williams06, williams11} 
constructed models of grain heating and destruction in non-radiative shocks to 
match {\it Spitzer} observations of young LMC SNRs. The spectra and intensities could be matched by 
models with fairly standard parameters, but the inferred pre-shock dust-to-gas ratio
in the ambient gas near the LMC remnants was typically 1/5 the average LMC 
value obtained from extinction 
studies \citep{WD01}.  \citet{arendt10} studied the
dust destruction in Puppis A, and the change in grain size distribution did not match 
that expected from parameters derived from X-ray spectra.  \citet{winkler13} find evidence
for a higher grain destruction rate in SN1006 than expected. 


\medskip
Most dust destruction in the ISM occurs in shocks at modest speeds in middle-aged SNRs simply because
they account for most of the volume swept out during SNR evolution.  A detailed
{\it Spitzer} study of grain destruction in the Cygnus Loop was carried out by \citet{sankrit}.
They obtained 24 $\mu$m and 70 $\mu$m images of a non-radiative shock in the northern Cygnus
Loop.  Relatively
faint optical and UV emission is produced in a narrow ionization zone just behind the shock,
and it provides several diagnostics for plasma temperatures \citep{cr78, ghavamian01}. 
Because the shock is non-radiative, there is no significant contribution of emission lines
to the IR spectrum \citep{williams11}.  The Cygnus Loop was chosen for this investigation because it is bright,
and because the small foreground E(B-V) means that it can be observed in the UV.  It is also
nearby, $<$ 640 pc \citep{blair09, salvesen09}, so that
the dust destruction zone is spatially resolved by X-ray and IR instruments.  \cite{sankrit}
selected a region where the shock parameters had been determined from H$\alpha$,
UV and X-ray observations \citep{ghavamian01, raymond03}.  

\medskip
The {\it Spitzer} 24 $\mu$m image is
shown along with H$\alpha$ and X-ray images in Figure 1.  Models
similar to those of \cite{borkowski06} were able to match the 
24 $\mu$m intensity falloff with distance,
the variation in the 24 $\mu$m/70 $\mu$m ratio and the IR to X-ray 
flux ratio.
The declines in the 24 $\mu$m/70 $\mu$m ratio and the IR to X-ray 
ratio clearly demonstrate destruction of dust, but there remain 
ambiguities involving the density and depth of the emitting region,
as well as the porosity of the grains.  \cite{sankrit} concluded that non-thermal
sputtering due to the motion of the grains through the plasma is required to match
the variation in the 24 $\mu$m/70 $\mu$m ratio.  That process is more important in the
400 \kms shocks of the Cygnus Loop than at the higher temperatures of the young SNRs
investigated by \citet{borkowski06}, \citet{williams06} and 
\citet{winkler13} because of the lower thermal speeds of protons and 
$\alpha$ particles behind the slower shock.  The best fit model of 
\cite{sankrit} has only half the dust-to-gas ratio expected for the diffuse ISM.

\medskip
In spite of the quality of the recent {\it Spitzer} observations, 
crucial questions about the heating and destruction of interstellar
dust in SNR shocks remain.  The inference that the dust-to-gas ratio
is only half the expected value could mean that either the derived
dust mass is too small due to incorrect heating rates or emissivities,
or else the destruction rate is underestimated.  In this paper we 
report the detection of emission in the C IV $\lambda$1550 doublet 
from carbon atoms liberated from grains behind the shock in the 
region observed by \cite{sankrit}.  Each neutral carbon atom liberated
from a grain is quickly ionized to C VI or C VII in the hot post-shock
gas, but during the time it spends in each ionization stage it can be
excited.  Thus each sputtered carbon atom emits about 30 photons in the
C IV doublet before it is ionized to C V.  We derive the rate at which carbon 
is liberated from grains and compare the observed intensities with 
model predictions.

\section{Observations and Data Reduction}

\begin{figure}
\epsscale{0.96}
\plotone{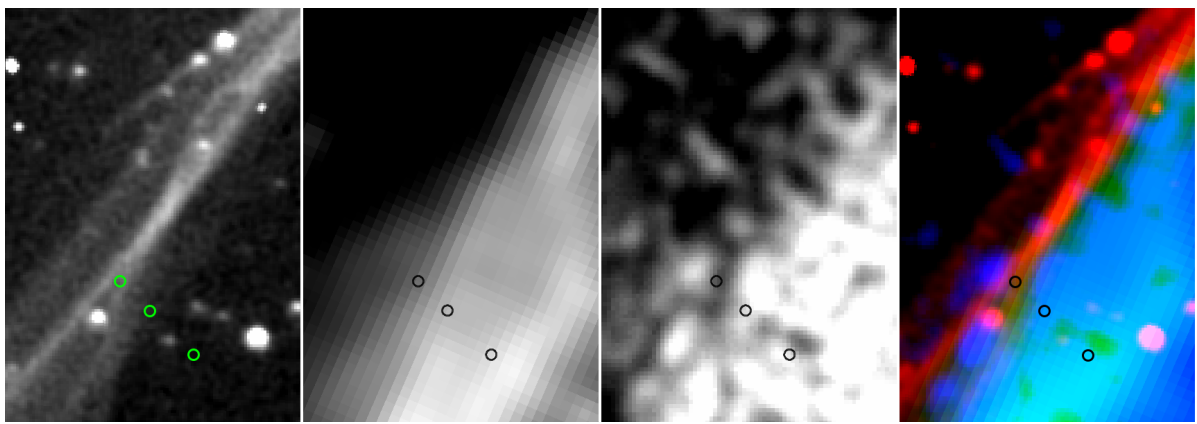}
\caption{ COS aperture positions overlaid on H$\alpha$,
{\it Spitzer} 24 $\mu$m images and Chandra X-ray images.  The right hand
panel shows Positions 1, 2 and 3 (left to right) overlaid on a 3 color superposition
of H$\alpha$ (red), 24 $\mu$m (green) and X-rays (blue).
The image scale is indicated by the 10$^\prime$$^\prime$
and 15$^\prime$$^\prime$ spacings between the COS aperture
positions.
}
\label{aperturepos}
\end{figure}

We observed three positions with the Cosmic Origins Spectrograph (COS) \citep{green} on
HST on 2012 April 25-26.  
Figure~\ref{aperturepos} shows the 3 observed positions overlaid on H$\alpha$,
Chandra X-ray and 24 $\mu$m images.  The 1.5" proper motion since the H$\alpha$
image was obtained was taken into account based on the value 
of 4.1" in 39 years measured by \cite{salvesen09}.  Table 1 shows
the positions and exposure times.  We used the G160M grating centered 
at 1577 \AA\/ with the PSA aperture covering the spectral ranges
1386 to 1559 \AA\/ and 1577 to 1751 \AA .   The positions
were chosen to be 0.4$^\prime$$^\prime$, 10$^\prime$$^\prime$, and 
25$^\prime$$^\prime$ behind the shock.  The first
position was intended to be slightly behind the shock position delineated
by H$\alpha$ because it takes a finite time, and therefore distance, to 
ionize carbon up to and through the C IV state.  For a
distance to the Cygnus Loop of 640 pc \citep{salvesen09}, a post-shock density
somewhat above 1 $\rm cm^{-3}$ \citep{raymond03} and a post-shock temperature
of about $2 \times 10^6$ K, that distance corresponds to about 0.4$^\prime$$^\prime$.
Unfortunately, the H$\alpha$ filament bifurcates at that position, and 
the COS aperture lies between the two segments.  From Figure~\ref{aperturepos}
it can be seen that the Position 1 aperture was centered on the H$\alpha$ filament about 
1.5$^\prime$$^\prime$ behind the brightness peak.

\begin{figure}
\epsscale{0.96}
\plotone{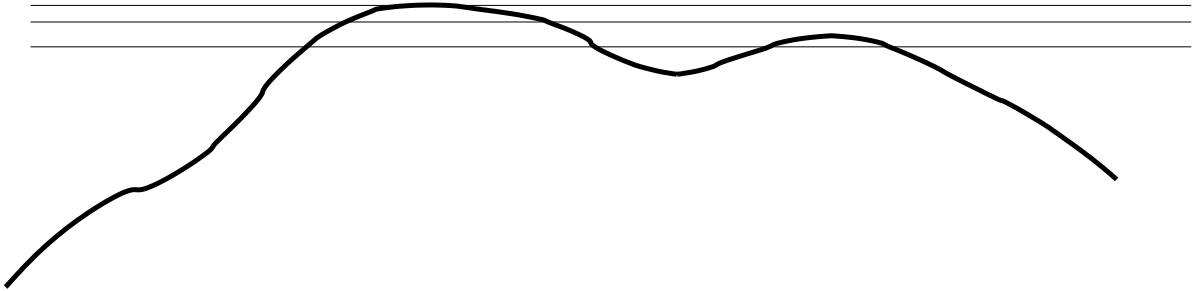}
\caption{Schematic diagram of the rippled shock surface and three lines of
sight at different distances behind the shock tangency.  The outermost
line of sight is tangent to the shock, giving a bright filament, while
the innermost line of sight would give a secondary brightness peak due
to near tangency with the second ripple.  The observer is located far to 
the left in this schematic.
}
\label{schematic}
\end{figure}

It should be noted, however, that appearances can be deceiving.  The SNR blast wave
is rippled as a result of velocity variations caused by density inhomogeneities
in the ambient ISM, and the H$\alpha$ filaments are actually tangencies between the line
of sight and the thin (unresolved) emitting region behind the shock \citep{hester87}.
A schematic diagram of the rippled shock surface and several lines of sight is
shown in Figure~\ref{schematic}.  The uppermost line of sight would correspond
to a bright filament, while the lowermost is close to a different tangent point,
so that it would appear bright.  Thus the apparent change in brightness as a function
of apparent distance behind the shock would contain a secondary brightness peak
unrelated to the outermost filament.

The data were processed with the standard COS pipeline except that we
fit a background plus emission lines to the spectrum rather than using the 
background-subtracted spectrum in order to get more a reliable estimate of the uncertainties.
The apparent continuum contains both the real hydrogen 2-photon continuum 
and the detector background, but we have not attempted to separate those
contributions.  The data were binned by 32 pixels for the fits. 

\begin{figure}
\epsscale{0.9}
\plotone{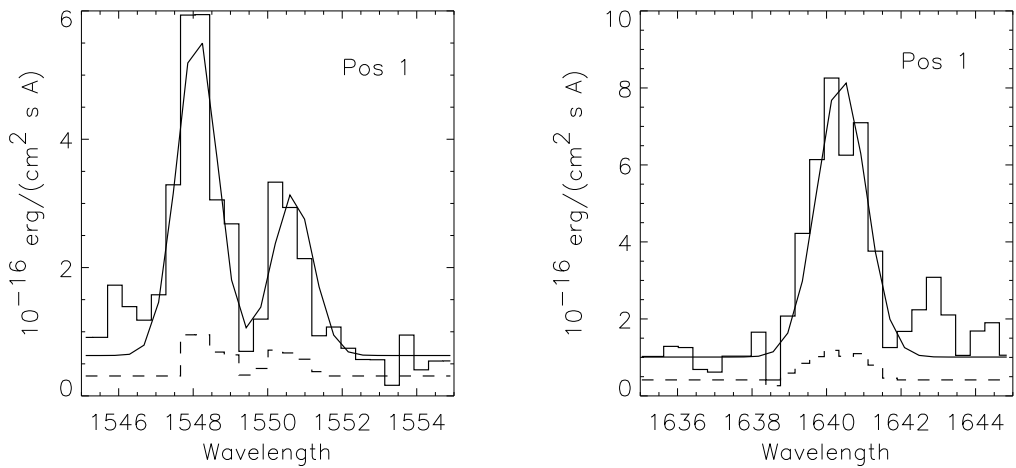}
\caption{ C IV doublet and He II $\lambda$1640
line at position 1.  The solid curve is the best
Gaussian fit, the solid histogram is the data,
and the dashed histogram shows the uncertainties.
}
\label{spct1}
\end{figure}

\begin{figure}
\epsscale{0.9}
\plotone{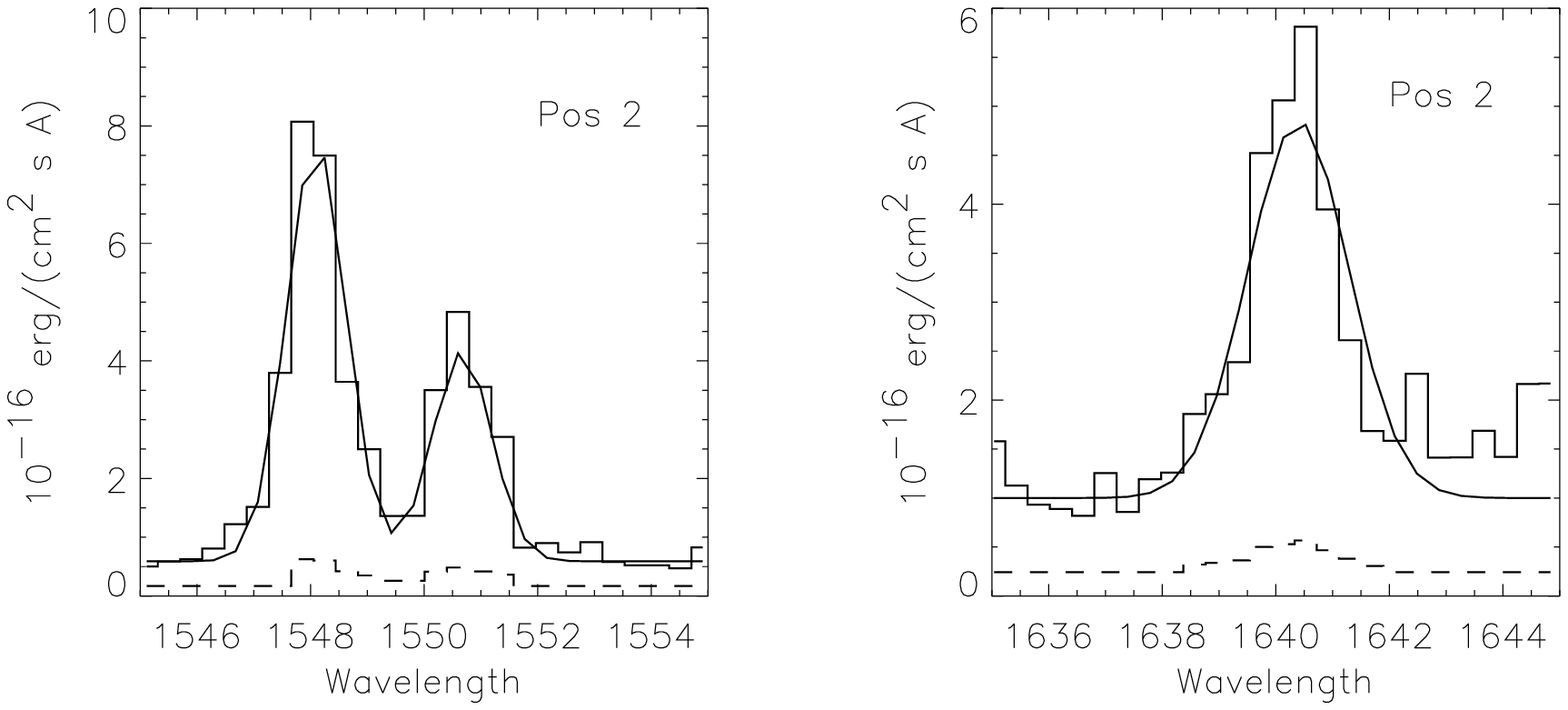}
\caption{ C IV doublet and He II $\lambda$1640
line at position 2.
}
\label{spct2}
\end{figure}

\begin{figure}
\epsscale{0.9}
\plotone{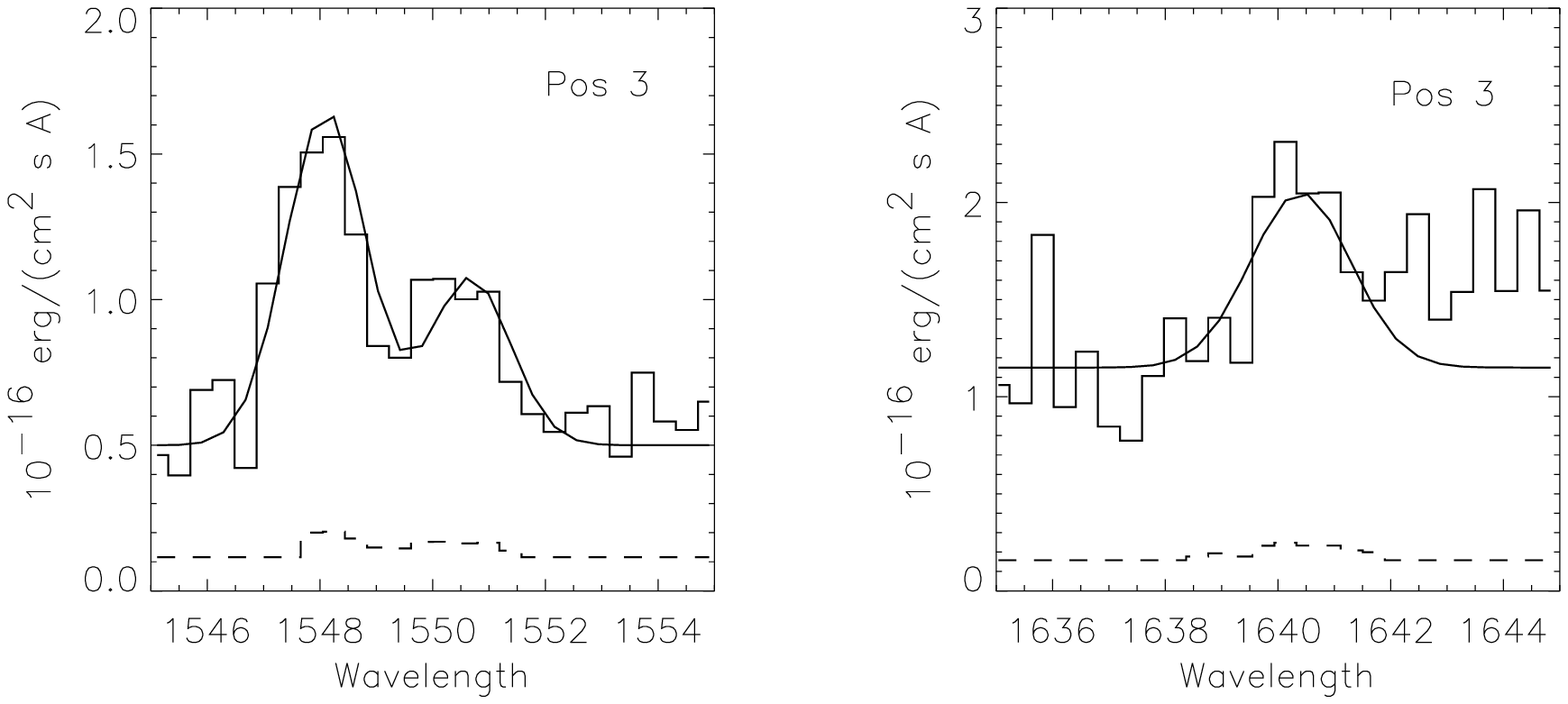}
\caption{ C IV doublet and He II $\lambda$1640
line at position 3.
}
\label{spct3}
\end{figure}

Figures~\ref{spct1} to~\ref{spct3} show the C IV and He II emission
lines at the three positions. The errors are based on the RMS deviations
from broad wavelength regions around the lines with an additional
contribution from the photon statistics of the lines.  For each position
we show the best Gaussian fits to the profiles, where the wavelength
separation of the C IV doublet is fixed at the laboratory value and
the intensity ratio is fixed at 2:1.  The Position 3 He II fit is
an exception, because the line is barely detected and the formal best fit has an 
unreasonably large width, so the fit shown assumes the Position 2 He II width.

The C IV $\lambda$1550 doublet and the He II $\lambda$1640 line were the only
features detected in the wavelength range sampled.  The next brightest features expected
are the Si IV $\lambda$$\lambda$ 1393,1402 doublet, the O III]
$\lambda$$\lambda$ 1664,1666 lines and the O IV] multiplet at
$\lambda$1400.  Those lines were not expected
to be detectable because of the low abundance
of Si, the high ionization rate of Si IV, and the small
excitation cross sections of the O III and O IV intercombination lines.
Thus, while these lines are easily detected in radiative shock
waves \citep{raymond88}, they are not seen in non-radiative shocks \citep{raymond95,
raymond03}.

The instrument profile of the COS aperture for an extended source is not
Gaussian.  Moreover, the Gaussian fits leave correlated residuals, which 
casts doubt on the accuracy of the fit.  Therefore, we measured intensities 
by simply integrating the fluxes above the background over the line profiles, 
and we use the integrated errors as estimates of the 1 $\sigma$ uncertainties. 
We do, however, use the Gaussian fits as the only reasonable measure of
the line widths.  The intensity ratio of the C IV doublet was fixed at its
intrinsic value of 2:1 for the fits to determine line widths.  The best fit 
line widths for the He II line were wider than those for the C IV lines by about
the amount expected if both helium and carbon are thermally broadened with a
temperature of about $1.5 \times 10^6$ K, but the uncertainties are larger
than the difference.

Table 2 shows the measured parameters
for the three positions.  The fluxes were corrected for a reddening
E(B-V) = 0.08 using the \cite{cardelli} galactic extinction function, as
adopted by \cite{raymond03}.  It is apparent that the C IV flux at Position 1
is smaller than that at Position 2.  This unexpected result is discussed below.

In a thin sheet of emitting gas seen edge-on, resonance scattering can
affect the observed intensities [e.g., \cite{long92}], reducing the intensity
ratio of the doublet from its intrinsic 2:1 value and reducing the total C IV
intensity.  \cite{raymond03} used Far Ultraviolet Spectroscopic Explorer
(FUSE) observations of the O VI doublet to exploit this effect to
constrain the shock parameters and geometry, and they showed that the optical depth in
O VI $\lambda$1032 is $\sim$ 1.  Considering the lower abundance
of carbon and the shorter ionization time of C IV, the optical
depth in the C IV $\lambda$1548 line should be $\sim$0.1.  However,
the FUSE spectra were acquired somewhat to the NW of our positions,
and they averaged over 20$^\prime$$^\prime$x4$^\prime$$^\prime$ regions,
so a larger optical depth at Position 1 is possible.  We performed Gaussian fits
with the I(1548)/I(1550) ratio unconstrained 
and found the ratio to be consistent with the optically thin ratio of 2:1.
The best fit ratio is actually slightly above 2, but lower ratios
corresponding to optical depths as large as
$\tau_{1548}=0.82$ are within the uncertainties, and that would reduce the 
total C IV flux at position 1 by at most 25\%.  Thus a small optical
depth is indicated by the I(1548)/I(1550) ratio.

We must also consider whether background emission from the Galaxy
makes a significant contribution to the observed fluxes. \cite{martin90}
measured C IV fluxes of 2700 to 5700 $\rm ph~cm^{-2}~s^{-1}~sr^{-1}$ at
high galactic latitudes, and they obtained only upper limits
somewhat below those values at low galactic latitudes where the Cygnus
Loop lies.  Even the highest Galactic background values are 35 times 
smaller than what we observe at Position 3, and we conclude that the measured
fluxes originate in the Cygnus Loop.

\section{Analysis}

We assume shock parameters based on optical, UV and X-ray studies of
a portion of the same H$\alpha$ filament located about 5.7$^\prime$
farther to the NW.  That region has a more complex H$\alpha$ morphology
due to several ripples of the shock surface [analogous to \cite{blair05}],
 so we chose a set of
positions along strip 1 of \cite{sankrit}.  \cite{ghavamian01} measured
a shock speed of 300-365 $\rm km~s^{-1}$ and a ratio of electron to
proton temperatures $T_e/T_p$ at the shock of 0.7-1.0 from the width
of the H$\alpha$ narrow component and the intensity ratio of the broad
and narrow components.  Subsequently, \cite{vanadelsberg} were unable
to match both the broad component width and the narrow-to-broad intensity
ratio, perhaps because of a contribution of a shock precursor to the
narrow component.  However, the electron temperature determined from
X-rays \citep{raymond03, salvesen09} supports the conclusion that 
$T_e$ is nearly equal to $T_p$, and in that case the broad component 
width indicates a shock speed at the upper end of the range given by 
\cite{ghavamian01}.  

UV observations of a section of the Balmer filament a few arcminutes
NW of our Position 1 by FUSE 
showed that the proton and oxygen kinetic temperatures were close to 
equilibration \citep{raymond03}.  The relative intensities of 
the C IV and He II lines in a UV spectrum from the Hopkins Ultraviolet
Telescope (HUT) indicated about half the solar carbon abundance, meaning that 
half the carbon entered the shock in the gas phase or in very small grains that were
vaporized within the HUT aperture, that is, within 5$^\prime$$^\prime$ 
of the shock \citep{raymond03}.  That paper also used 
the optical depths in the O VI lines to estimate a depth along the 
line-of-sight of 0.7-1.5 pc, pre-shock density of 0.3-0.5 $\rm cm^{-3}$ 
and a pre-shock neutral fraction of about 0.5.  \cite{salvesen09} measured
proper motions of filaments along the northern Cygnus Loop.  Their
filament 6 coincides with our Position 1, and the proper motion is
0.105"/yr or 333 $\rm km~s^{-1}$ at 640 pc.  \cite{katsuda} analyzed Chandra observations
of the NE region of the Cygnus Loop, and our Position 1 is located near the
SE end of their Area 1.  The emission measure they derive is compatible with
a pre-shock density of 0.5 $\rm cm^{-3}$ and a line-of-sight depth of
1.5 to 2 pc.  However, their derived electron temperature of about 0.27 keV
is about twice what one would expect from the 333 $\rm km~s^{-1}$ shock
speed given by the proper motion and the 640 pc upper limit to the distance
of \citet{blair09}. 

To interpret the line fluxes, we need to know how many photons each
atom emits before it is ionized.  Since the post-shock temperature
is far above the temperature where C IV and He II are found in ionization
equilibrium (log T = 5.0 and 4.7, respectively), each atom survives for
a time $\tau_{ion} = 1/(n_e q_i)$ and it is excited at a rate $n_e q_{ex}$.
Therefore, it emits on average $q_{ex}/q_i$ photons, where $q_{ex}$ is the 
excitation rate and $q_i$ is the ionization rate, before it is ionized.  
Using Version 6 of CHIANTI \citep{dere09}, specifically the He II excitation computed by
Connor Ballance for that database and the C IV excitation rate from \cite{griffin},
with the ionization rates of 
\cite{dere07}, we find that each C atom emits 31 $\lambda$1550 photons, while each He 
atom emits 0.078 $\lambda$1640 photons.  Thus for solar abundances \citep{asplund}, 
the C: He ratio of 0.0032 implies an 
intensity ratio I(C IV)/I(He II) = 1.33 (in $\rm erg~cm^{-2}~s^{-1}$).
We will also use the similar number for hydrogen;  each neutral H atom
passing through the shock produces 0.25 H$\alpha$ photons \citep{ckr}.

The H$\alpha$ image shown in Figure~\ref{aperturepos} was obtained in 1999 at the
1.2 m telescope at the Fred Lawrence Whipple Observatory.  It was calibrated based on the 
optical spectrum at a nearby position, and H$\alpha$ fluxes in several apertures
are given in \cite{raymond03}.  We determined the H$\alpha$ fluxes within the
COS apertures at the three positions, and they are shown in Table 2.
The intensity ratios can be combined with the numbers of photons per
atom to infer the neutral fraction of hydrogen entering the shock. We
asssume that helium is entirely neutral or singly ionized. Substantial
numbers of He I $\lambda$584 and He II$\lambda$ 304 photons can ionize H
and He I, but the shock produces relatively few photons above 54.4 eV,
and the photoionization cross section at those energies is relatively small.
We use the ratio of $\lambda$1640 intensity to H$\alpha$ at Position 1 to derive
a neutral fraction of 0.11
with an uncertainty of a factor of 1.8, including a 32\% measurement uncertainty
in the $\lambda$1640 intensity (2-$\sigma$) and uncertainty estimates in reddening correction
and H$\alpha$ calibration.  We therefore estimate a hydrogen neutral fraction of 0.06 to 0.20.
That is smaller than the more model-dependent estimate of \cite{raymond03}, but in keeping
with the upper limit of 0.2 from the limit on the ratio of He II $\lambda$ to H$\alpha$
\citep{ghavamian01}.

\subsection{Gas phase and sputtered carbon contributions to C IV}
\label{gasphase}
 
The observed intensities are the sum of emission from C atoms liberated
from dust grains downstream of the shock and emission from near the shock 
as it curves around the Cygnus Loop, projected onto the  line of sight.  
On the other hand, He is very quickly ionized, and since it 
is not depleted onto grains, the $\lambda$1640 emission is produced only 
at the shock front.  Based on the temperatures and densities above, it 
originates within 1$^\prime$$^\prime$ of the shock.  The aperture
at Position 1 includes both ``gas phase" C IV emission and emission from 
carbon sputtered from dust grains that passed through parts of the shock 
that appear ahead of Position 1 in projection.  Therefore, we take the Position 1
ratio of I(C IV)/I(He II) = 1.1 to be an upper limit to the ratio due
to emission at the shock from carbon that passes through the shock in
the gas phase.  This includes emission from PAHs and very small grains that
are vaporized near the shock \citep{micelotta}.  Comparison of the value of 1.1
with the theoretical value of 1.33 indicates that at most 80\% of the carbon is
in the gas phase or PAHs at the shock.  
HUT measured a C IV: He II ratio of 0.73 for a section of this filament
farther to the NW.  The 
10$^\prime$$^\prime$ HUT wide aperture was placed along the filament, so it 
includes C IV emission from carbon that is vaporized from  
grains within 5$^\prime$$^\prime$ of the shock.  This gives a more
stringent limit of 0.45 for the fraction of carbon entering the shock
in the gas phase or PAHs and it suggests a significant amount of sputtering
within 5$^\prime$$^\prime$ of the shock.  The HUT upper limit is comparable to 
the estimated dust-to-gas ratio of about one half the Galactic value from
\cite{sankrit}.

We assume that the ratio of C IV to He II produced at the shock is constant
and use it to subtract off the contribution to the C IV emission from the 
shocks projected onto the line of sight at Positions 2 and 3. 
Taking the C IV:He II ratio at the shock to be $<1.1$, we find that
the C IV emission from carbon liberated from dust is 16 to 27 and
3.3 to 5.5$\times 10^{-16}~\rm erg~cm^{-2}~s^{-1}$ at positions 2 and
3, respectively.  Those fluxes imply that carbon is being sputtered
from grains at rates of 4.6 to $7.7 \times 10^5$ atoms $\rm cm^{-2}~s^{-1}$
at Position 2 and 0.9 to $1.5 \times 10^5$ atoms $\rm cm^{-2}~s^{-1}$
at Position 3.

\subsection{Line widths}

The measured widths of the $\lambda$1640 profiles are larger than
those of the C IV lines as expected if their kinetic temperatures
are equal, but the widths are equal within the uncertainties.  The
thermal width of the helium line is 141 $\rm km~s^{-1}$ (FWHM) at 
$1.7 \times 10^6$ K expected for the 350 \kms shock speed obtained
by \citet{ghavamian01} from the H$\alpha$ profile, compared with 
82 $\rm km~s^{-1}$ for carbon.  The
COS line profile for diffuse emission that fills the aperture is
not exactly known beyond the statement that the width is about 
200 $\rm km~s^{-1}$ \citep{france}, so the observed line widths are consistent 
with thermal broadening and equal carbon and helium and hydrogen 
kinetic temperatures.  This is in contrast with shocks in
the solar wind \citep{korreck07, zimbardo} and faster SNR shocks
\citep{raymond95, korreck04}, where 
more massive ions have much higher temperatures, but the uncertainties
permit a wide range of kinetic temperatures.  Moreover,
the larger grains slow down very gradually in the shocked plasma 
due to Coulomb collisions, though they may gyrate about the magnetic field
and move with the bulk flow \citep{dwek96,sankrit}.  Carbon atoms 
sputtered from these grains will initially move at a high speed, potentially 
giving a line width comparable to the H$\alpha$ line width, as we
discuss below.

\section{Physical Models}

\cite{sankrit} found a dust-to-gas ratio about half the typical galactic
value, and we used their model to predict the C IV emission, using the
fraction of dust remaining as a function of distance behind the shock to
estimate the sputtering rate and assuming 31 photons per C atom.  
We measure a higher C IV intensity than expected from a simple plane-parallel
model, suggesting
that emission from gas phase carbon makes a significant contribution
or that the simple plane parallel model is not adequate. 
We therefore make more detailed models of the destruction of grains and
the emission from sputtered carbon, and we use those models with the
shock geometry inferred from the H$\alpha$ image to predict the C IV
brightness for comparison with the observed values.

\subsection{Grain Destruction Models}

\cite{sankrit} computed models of grain destruction and IR emission
to match {\it Spitzer} observations of this part of the Cygnus Loop, and
we have used a model close to their lowest temperature model, which matches
the shock speed determined from the proper motion, to predict the C IV
emission from carbon sputtered from dust.  The model, which is described
more fully in \cite{williams}, includes the enhanced sputtering due to 
the motion of grains through the hot plasma.  
Sputtering of a dust grain in a hot plasma is a function of both the 
temperature (energy per collision) and density (frequency of collisions) 
of the gas. We include sputtering by both protons and alpha particles, 
assuming cosmic abundances, such that n$_{\alpha}$ = 0.1n$_{p}$. We use 
sputtering rates from \citet{nozawa06}, augmented by calculations of an 
enhancement in sputtering yields for small grains by \citet{jurac}. We 
use the pre-shock grain size distributions of \citet{WD01} and 
calculate the sputtering for grain sizes from 1 nm to 1 $\mu$m as a 
function of the sputtering timescale, $\tau$, defined as the integral of 
the proton density over the time since the gas was shocked.  As a comparison,
we also computed sputtering rates for model BARE-GR-S of \citet{zubko}, and
found rates that were over twice as large near the shock and about 50\%
higher for $\tau$ corresponding to our Position
2 and 30\% higher for Position 3.  

The total mass 
in grains is calculated by integrating over the grain-size distribution, 
which changes as a function of $\tau$ due to sputtering. Relative motions 
between the dust grains and the hot gas, which result in ``non-thermal" 
sputtering, are included by solving the coupled differential equations for 
grain radius and velocity given in \citet{ds79}. Dust grains enter the 
shock with a velocity of 3v$_{s}$/4 relative to the downstream plasma, and 
slow down due to collisions with the ambient gas. Initially, sputtering is 
a combination of both thermal and non-thermal effects, with the non-thermal 
effects going to zero as gas drag and to a lesser extent Coulomb drag slow 
the grains with respect to the gas.  Just behind the shock, the motion of
the grains through the plasma approximately doubles the sputtering rate.  By
the time the gas reaches our Positions 2 and 3, the grains smaller than about
0.01 $\mu$m have slowed enough that sputtering rates approach the thermal 
value, while grains about 0.1 $\mu$m still experience the enhanced rate.
Based on the results
of \cite{ghavamian01} and \cite{raymond03} we assume
equal electron and ion temperatures of 160 eV, which corresponds to
a shock speed of 366 \kms. \cite{katsuda} find a higher electron 
temperature of about 270 eV for regions to the NW of our positions,
but we adopt the shock speed based on measured proper motions
\citep{salvesen09}. 

Figure~\ref{models} shows the predicted dust
destruction rates and fractions of dust remaining as a function of
$\tau$ behind the shock.  Silicates behave somewhat differently
than carbonaceous grains, and we show the silicate curves for comparison.
For a post-shock density of 2 $\rm cm^{-3}$,
the shock proper motion of 0.105$^\prime$$^\prime$ per year and an
assumed compression of a factor of 4 by the shock (so that the
post-shock gas moves at 0.0265$^\prime$$^\prime$ per year relative to the
shock), Positions 2 and 3 correspond to $2.4\times 10^{10}$ and
$5.9\times 10^{10}~\rm cm^{-3}~s$, respectively.  The model assumes
that 25\% of the carbon is initially in atomic or molecular form
or in PAHs that are destroyed very rapidly close to the shock.

Figure~\ref{grainveloc} shows how grains are decelerated behind the shock by the
gas drag (Baines et al. 1965; Draine \& Salpeter 1979), for grains with
initial (preshock) radii of 0.005, 0.01, 0.02, 0.04, and 0.1 $\mu$m. (We
did not include the Coulomb drag in these calculations, as it is
generally less important than the gas drag in hot X-ray emitting
plasmas due to the small values of charge on the grains). For large grains, 
deceleration is modest on temporal scales of
interest in this paper because the slowing-down time (drag time) is
long. As the drag time scales linearly with the grain density,
carbonaceous grains are decelerated more rapidly than silicate grains.
For small grains, both sputtering and gas drag become important, and
their combined effects lead to rapid deceleration. This is particularly
pronounced for silicate grains because their sputtering yields
are larger than for carbonaceous grains (Nozawa et al. 2006). For
example, silicate grains with the initial grain size of 0.005 $\mu$m
quickly slow down and are completely sputtered away at $\tau = 3.5
\times 10^{10}$ cm$^{-3}$ s.

Figure~\ref{grain_destruction} shows the mass fraction in grains as a function of
velocity relative to the post-shock plasma.  Again, solid lines depict carbonaceous grains
with initial radii of 0.005, 0.01, 0.02, 0.04, and 0.1 $\mu$m from bottom to top.
Much of the carbon liberated from grains comes from the smaller ones, so much of
the carbon is produced when grains have slowed by about a factor of 2 relative to 
the gas.  This will affect the width of the C IV lines from sputtered carbon.
If the shock is a parallel shock, the dust is not initially compressed by a factor
of 4 along with the gas, but it is compressed as the velocity decreases.  That could
affect the spatial distribution of C IV emission and IR intensity behind the shock.
Figure~\ref{grain_destruction} also indicates the velocities of the grains from which the carbon
is sputtered.  A preliminary calculation of the line profile at Position 2 assuming
a perpendicular shock and a magnetic field in the plane of the sky shows an emission
plateau 540 \kms wide (FWZI), with a central component of about 200 \kms (FWHM).  For other
magnetic geometries, those widths would be multiplied by the cosine of the angle
between the shock normal and the magnetic field and by the sine of the angle
between the magnetic field and the line of sight.  However, the motion of particles
along the field relative to the plasma will affect the velocity distribution at a given
location.  The measured FWHM at Position 2
is just within the uncertainties of the predicted central component width.  The
predicted profile at Position 3 consists mostly of the broad plateau, which is compatible
with the 460 \kms upper limit to the measured velocity width at that
position.

\subsection{Comparison to observations}

\begin{figure}
\epsscale{0.9}
\plotone{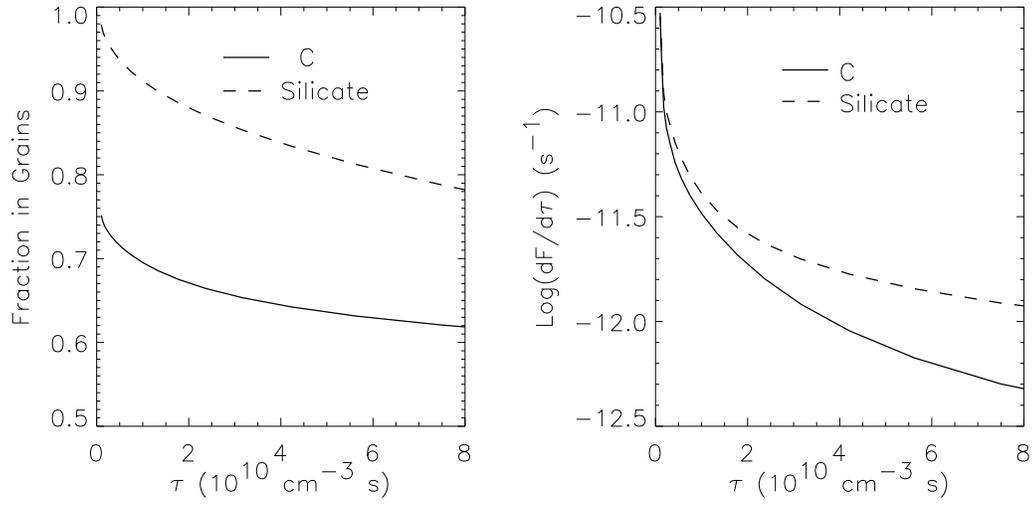}
\caption{Left panel; predictions for the fractions of carbon
and silicates remaining in grains as a function of
$\tau = n_e t$ according to the model described in section
4.1.  Right panel; rates at which carbon and silicates are
sputtered from grains as a function of $\tau$.
}
\label{models}
\end{figure}

\begin{figure}
\epsscale{0.9}
\plotone{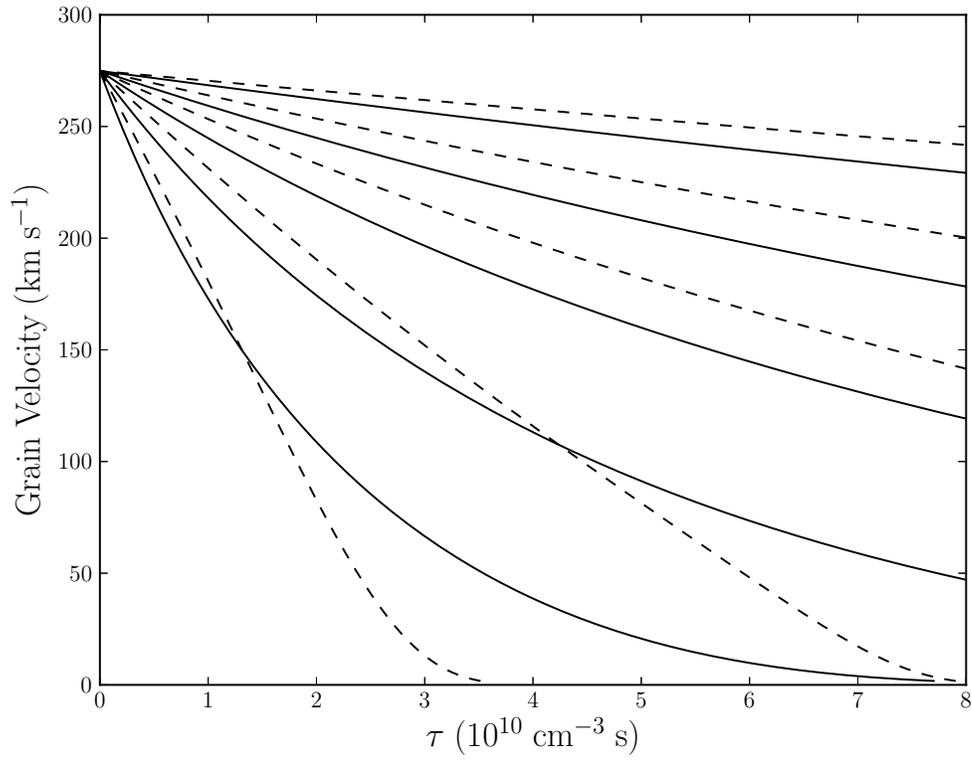}
\caption{Grain velocities relative to the shocked plasma as a function of
$\tau$ for grain sizes 0.1, 0.04, 0.02, 0.01, and 0.005$\mu$m (top to bottom).
Solid curves are for carbonaceous grains, and dashed curves for silicates.
}
\label{grainveloc}
\end{figure}

\begin{figure}
\epsscale{0.9}
\plotone{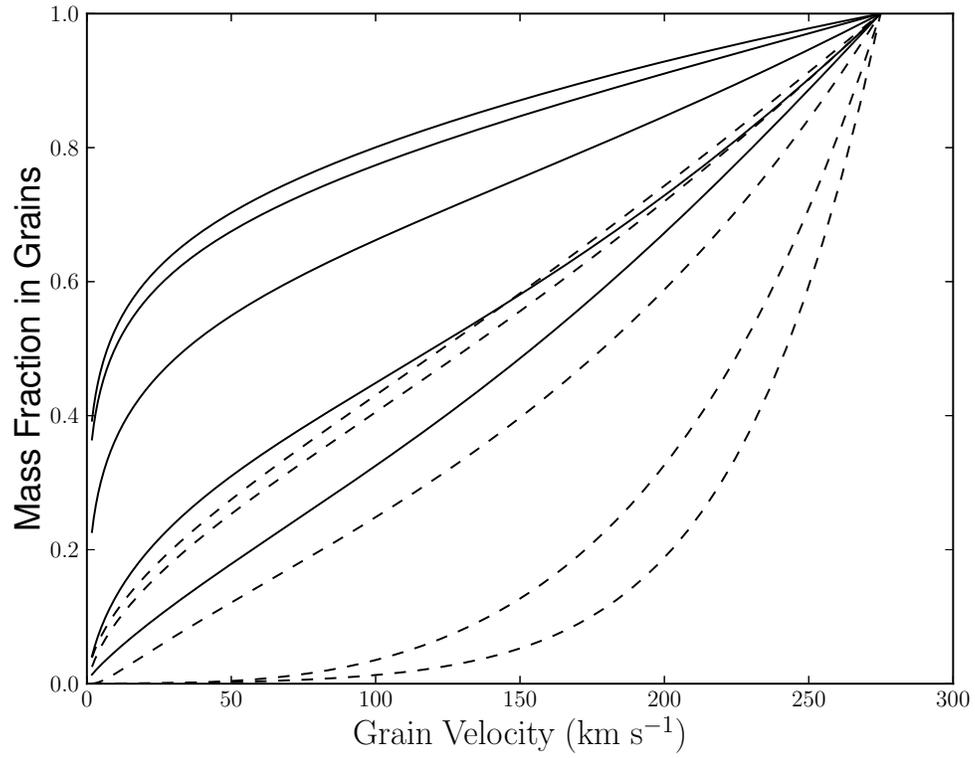}
\caption{Mass fraction in grains as a function of velocity relative to the shocked plasma 
for grain sizes 0.1, 0.04, 0.02, 0.01, and 0.005$\mu$m (top to bottom).  Solid lines
pertain to carbonaceous grains, and dashed lines show silicate grains for comparison.
}
\label{grain_destruction}
\end{figure}

As can be seen in Figure~\ref{aperturepos}, the shock is not a simple
planar sheet conveniently oriented along our line of sight.  Rather,
it is a rippled sheet that appears bright where it is tangent to
our line of sight \citep{hester87}.  Therefore, we cannot simply
compare the model prediction in Figure~\ref{models} with the observed
fluxes without considering the geometrical structure of the shock front.  
Indeed, the sputtering rate $dF/d\tau$ from Figure~\ref{models} drops
steeply, and for $\tau$
about $5 \times 10^{10}~\rm cm^{-3}~s$ it is very low, so that multiplying that
value by the column density of carbon and the photon yield per atom yields 
a C IV flux below that observed. 

\begin{figure}
\epsscale{0.9}
\plotone{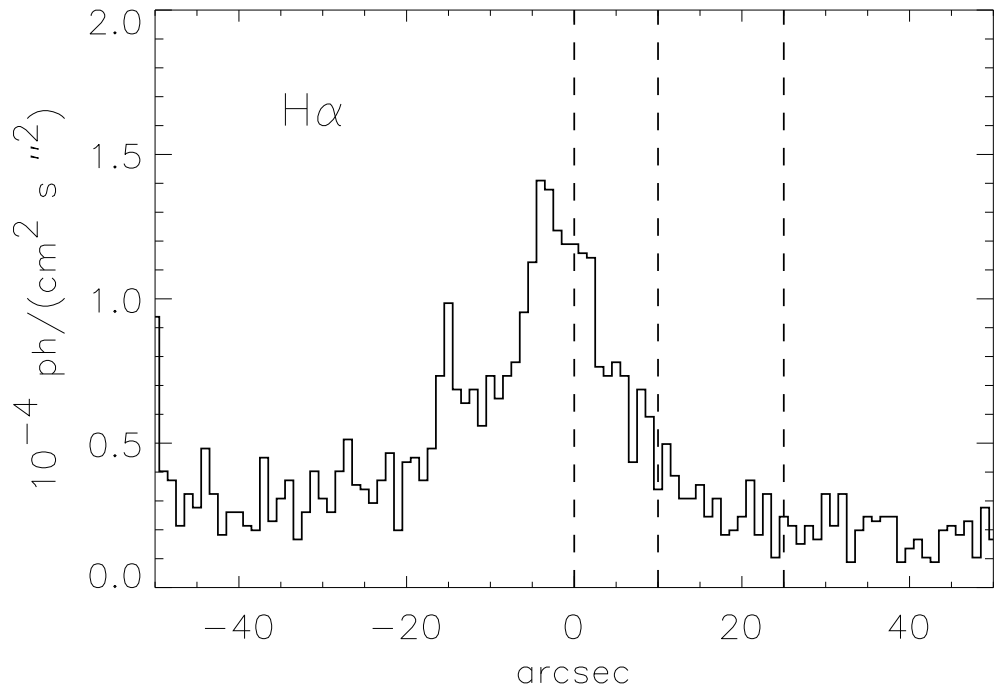}
\caption{H$\alpha$ surface brightness along a line through
Positions 1, 2 and 3 from a 2.8$^\prime$$^\prime$ (3 pixel) wide
average from the image in Figure 1. Positions 1, 2 and 3 are located
at x-values 0, 10 and 25.  There is a star at -50.  
}
\label{halphacut}
\end{figure}

To describe the shape of the shock front, we use the H$\alpha$ image
shown in Figure~\ref{aperturepos}.  The intensities from a 3 pixel
wide average along the line connecting Positions 1, 2 and 3 are shown
in Figure~\ref{halphacut}.  It was shown above that the neutral fraction
in the pre-shock gas is 0.06 to 0.2, and each H atom produces 0.25 H$\alpha$
photons just after it passes through the shock, so the H$\alpha$ brightness
is directly related to the flux of particles through the shock at each
pixel.  We next assume that the ratio of carbon to hydrogen is $3 \times 10^{-4}$
by number \citep{asplund} with 75\% in dust and use the model shown in Figure~\ref{models}
to compute the C IV intensity at each pixel along the cut through
Positions 1, 2 and 3 assuming a post-shock density of 2 $\rm cm^{-3}$. 
Note that the intensity has two components. First,
there is C IV produced immediately behind the shock from carbon that
was in the gas phase or PAHs, which is proportional to the local H$\alpha$ brightness.
Second, there is the C IV
from carbon sputtered from grains that passed through the shock in pixels
farther toward the outside of the remnant.  We assume that the H$\alpha$ is formed
at the shock front, and its brightness indicates the mass flux through the shock
at each position as shown in Figure ~\ref{halphacut}.  We then use the emission 
as a function of spatial offset from the shock derived from Figure~\ref{models}
to compute the C IV emission from gas at all downstream pixels and sum the
contributions from the shocks the positions along Figure~\ref{halphacut}.

\begin{figure}
\epsscale{0.9}
\plotone{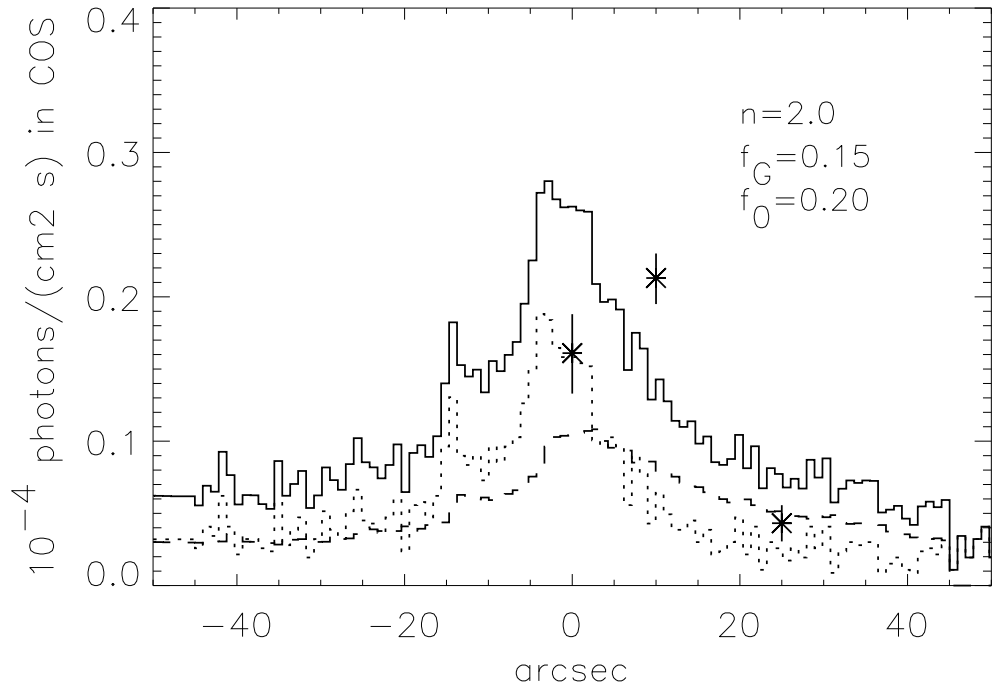}
\caption{Predicted C IV surface brightness due to carbon that
is in the gas phase at the shock (short dashed), carbon sputtered from
grains (long dashed) and the total (solid).  COS observations at
positions 1, 2 and 3 are shown with 2$\sigma$ error bars.
}
\label{civcut}
\end{figure}

Figure~\ref{civcut} shows the predicted C IV intensities along with the
observed values, where we have assumed a pre-shock neutral fraction of 0.2, at
the upper end of the range determined above.  The predictions lie above the 
observations at Positions 1 and 3, but below the observations at Position 2.
Note that the 
predicted rate of liberation of carbon from grains, $dF/d\tau$ in Figure~\ref{models},
drops steeply with $\tau$, and much of the emission at Position 3 arises
from gas that was shocked relatively close to Position 3 (in projection) rather than gas 
that passed through the shock seen as the H$\alpha$ filament.  

Overall, the approximate agreement between the observed and predicted fluxes is
encouraging considering the uncertainties in the gas-phase abundances at the shock,
the sputtering rate, the grain size distribution, the post-shock temperature
and the pre-shock neutral fraction.  If some of the H$\alpha$ arises in a shock
precursor \citep{hrb94,raymond11}, the mass flux through the shock and therefore
the predicted C IV emission would be overestimated.  On the other hand, if the 
pre-shock neutral fraction is overestimated, the mass flux is underestimated
and the predicted C IV is underestimated.  If the post-shock temperature we
have assumed is underestimated, the sputtering rate is also underestimated.
Another uncertainty is that some of the H$\alpha$ emission can arise from
the photoionization precursor.  Though this is absent behind the shock in
the plane parallel case, the curvature of the SNR blastwave means that some
of the precursor emission will be seen in projection behind the shock, leading
to an overprediction of the C IV intensity.
We have no way to resolve these ambiguities, but conclude that a model of grain
destruction with current sputtering rates, combined with plausible parameters
for the shock, predicts a level of C IV emission from C atoms liberated from grains in
rough agreement with observations.
 
However, the discrepancy between Position 2 and the other 2 positions remains.
There are several possible explanations.

{\it Neutral fraction variations:} We have assumed that the hydrogen
neutral fraction is constant throughout the relevant part of the ISM.  
In a region where the neutral fraction is larger, a given H$\alpha$
brightness would translate into a lower mass flux
than the value used in the model, and the C IV brightness would be smaller.
We derived the neutral fraction from the Position 1 observation,
so this explanation would require that the neutral fraction changes
between Positions 1 and 2.

\begin{figure}
\epsscale{0.9}
\plotone{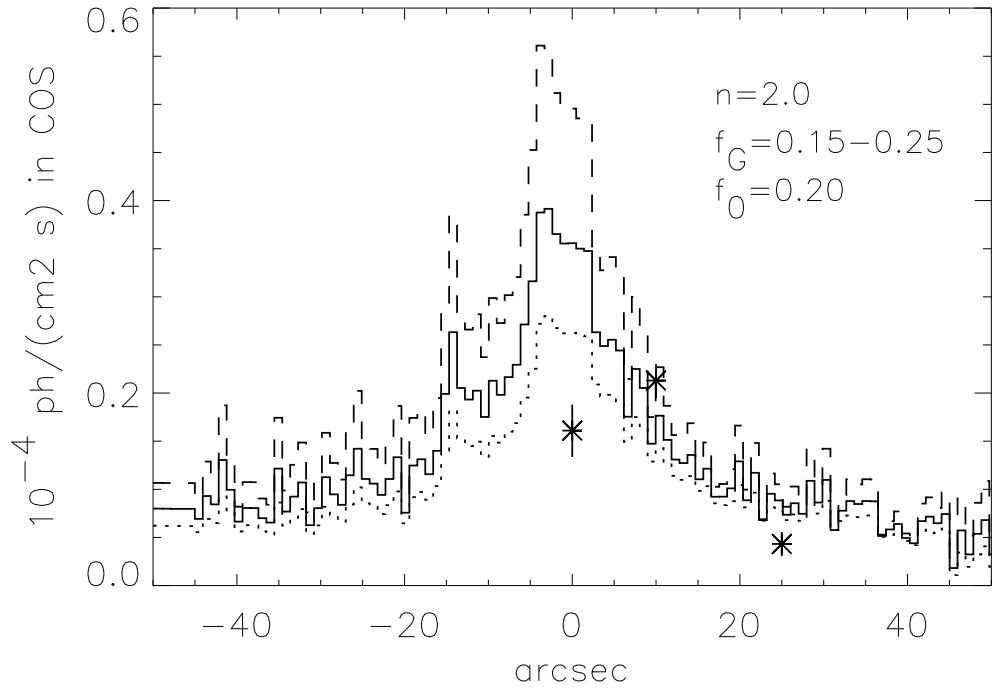}
\caption{Predicted C IV surface brightness due to carbon for gas
phase fractions of carbon at the shock of 0.25 (solid), 0.40 (dashed)
and 0.15 (dotted).  COS observations at
positions 1, 2 and 3 are shown with 2$\sigma$ error bars.
}
\label{gasfrac}
\end{figure}

{\it Gas phase carbon variations:}  The model assumes that the fraction
of carbon in the gas phase at the shock is 0.25 everywhere.  The value quite
likely varies by a factor of 2 in the ISM \citep{jenkins, sofia}. 
Figure~\ref{gasfrac} shows that reducing the gas phase fraction to about 
10\% would bring the Position 1 C IV flux into agreement but then the model
underpredicts the intensity at Position 2. 

{\it Sputtering rate:} The sputtering rate is poorly known \citep{nozawa06},
and it scales with the post-shock density, which our models assume 
to be 2 $\rm cm^{-3}$.  Increasing the sputtering rate would increase the 
C IV intensity from sputtered carbon, especially in pixels just behind the 
shock.  A combination of smaller fraction of carbon in the
gas phase and higher sputtering rate might in principle decrease the level
of disagreement, but it would not give a higher C IV intensity at
Position 2 than at Position 1.  Sputtering rates computed with the \citet{zubko}
size distribution would be higher everywhere, and they would not help
resolve the discrepancy.

{\it Optical depth:} If resonant scattering in the edge-on sheet of gas
just behind the shock reduces the C IV intensity at Position 1 by a factor of 
2, but does not affect Positions 2 or 3, that would solve the problem.  However,
that would require an optical depth of 3.5 in the $\lambda$1548 line,
which would imply a C IV doublet ratio of only 1.2, which is not
compatible with the observed spectrum.

{\it ISM inhomogeneity:} The $H\alpha$ intensity may not be an adequate
proxy for the flux of material through the shock.  The IR surface brightness
from {\it Spitzer} peaks perhaps 30" behind the bright H$\alpha$ filament,
which is not consistent with either a planar shock picture or with the
convolution of the H$\alpha$ brightness with the sputtering model.
The bright IR emission could indicate that the shock seen in projection
about 20" behind the H$\alpha$ filament is now passing through
a region with very low hydrogen neutral fraction, or it could be that the
shock passed through a dense or dusty clump about 1000 years ago.  A similar
variation in density along the shock path was noted by \citet{winkler13}
in SN1006.

Overall, while we have been able to detect the C IV emission from carbon sputtered
from grains, it is likely that projection effects (caused by variations in 
density, changes in neutral fraction, variable dust properties or some combination
of these) limit our ability to provide a definitive test of grain destruction
models with these data.

\section{Summary}

A simple, model-independent estimate of the C IV emission from carbon atoms
sputtered from dust grains behind the shock was made by assuming that the
ratio of C IV to He II emission from carbon in the gas phase at the shock
is constant over the region observed.  An attempt to model the
emission in more detail by using the H$\alpha$ intensity to map out the
geometrical structure of the shock produced general agreement to about a 
factor of 1.5, but a significant discrepancy among the positions remained.  In particular,
the models cannot explain why the C IV is brighter at Position 2 than at
Position 1.  We considered several explanations for this discrepancy, but 
the most likely is that the properties of the ISM, in particular 
density, neutral fraction or dust properties, vary over parsec scales.
Overall, we find that the C IV emission from carbon sputtered
from grains is compatible with the sputtering rate and grain size
distribution assumed by \cite{sankrit}, but preferably with a smaller
fraction of carbon in the gas phase and PAHs than assumed by those models.

The complexity of the structure along the line of sight prevents us from
deriving stronger constraints at present.
Further progress could be achieved by 1) observing a simpler shock structure
in the Cygnus Loop or another SNR, 2) resolving the discrepancy between shock 
speed derived from proper motions and post-shock temperature from the X-ray 
spectra, 3) sorting out the geometrical complexity of this part of the
Cygnus Loop, for instance by obtaining more observations of the He II line
and a higher resolution H$\alpha$ image, or 5) obtaining deeper X-ray spectra
to better determine the temperature, density and elemental abundances of
the shocked plasma.

\bigskip
This work was performed under grant HST-GO-12885 to the Smithsonian
Astrophysical Observatory.  P.G was supported under grant HST-GO-12545.08,
and T.J.G. acknowledges support under NASA contract NAS8-03060.
  
{\it Facilities:} \facility{HST (COS)} \facility{FLWO:1.2m}


\begin{table}
\begin{center}

\centerline{Table 1}
\vspace{2mm}
\centerline{COS Observations}
\vspace{2mm}
\begin{tabular}{ l c c c c }
\hline
\hline
Position   &   RA(2000)   & Dec(2000)  &  Dist. from Shock    & Exp. Time  \\
\hline
1  &  20 54 43.611 &  32 16 03.53 &  0.43$^\prime$$^\prime$ &  2500   \\
2  &  20 54 43.055 &  32 15 56.46 &    10$^\prime$$^\prime$ &  7801   \\
3  &  20 54 42.221 &  32 15 45.85 &    25$^\prime$$^\prime$ & 14501    \\

\hline
\end{tabular}
\end{center}
\end{table}

\begin{table}
\begin{center}


\centerline{Table 2}
\vspace{2mm}
\centerline{C IV $\lambda$1550 and He II $\lambda$1640 Fluxes and Widths}
\vspace{2mm}
\begin{tabular}{ l c c c c c c c}
\hline
\hline
\multicolumn{1}{c}{} & \multicolumn{4}{c}{Observed} & \multicolumn{3}{c}{Dereddened} \\

Position   &   F$_{1550}^a$   & w$_{1550}^b$  &  F$_{1640}^a$    & w$_{1640}^b$ & I$_{1550}^a$ & I$_{1640}^a$ & I$_{H\alpha }^c$ \\
\hline
1  &  11.2$\pm$0.96 & $247_{-68}^{+110}$ & 10.1$\pm$1.6  & $280_{-70}^{+104}$ & 20.7  & 18.3  & 17.4 \\
2  &  14.8$\pm$0.59 & $237_{-34}^{+38}$  &  5.5$\pm$0.94 & $384_{-134}^{+195}$ & 27.3  &  9.95 & 7.36 \\
3  &  3.0$\pm$0.44  & $324_{-92}^{+135}$  &  1.1$\pm$0.69 & $1340_{-1020}^{+240}$ &  5.54 &  1.99 & 2.79 \\

\hline
\end{tabular}
\end{center}
$^a$ $10^{-16}~erg~cm^{-2}~s^{-1}$

$^b$ FWHM uncorrected for instrument profile: km s$^{-1}$

$^c$ from H$\alpha$ image from Mt. Hopkins 1.2-m telescope

\end{table}

\newpage

\end{document}